\documentstyle[pra,aps,graphicx,epsfig]{revtex}
\begin{document}
\draft
\twocolumn[\hsize\textwidth\columnwidth\hsize\csname @twocolumnfalse\endcsname
\author{F. Casagrande, A. Ferraro, A. Lulli, R. Bonifacio}
\title{Measurement of the micromaser linewidth}
\address{INFM - Dipartimento di Fisica, Universit\`a di Milano, Via Celoria 16, 20133 Milano, Italy}

\author{E. Solano$^{1,2}$ and H. Walther$^1$}
\title{Measurement of the phase diffusion dynamics in the micromaser}
\address{$^1$Max-Planck Institut f\"ur Quantenoptik, Hans-Kopfermann Strasse 1, 85748 Garching, Germany \\
$^{2}$Secci\'{o}n F\'{\i}sica, Departamento de Ciencias,
Pontificia Universidad Cat\'{o}lica del Per\'{u}, Apartado 1761,
Lima, Peru}

\date{\today}
\maketitle

\begin{abstract}
We propose a realistic scheme for measuring the micromaser
linewidth by monitoring the phase diffusion dynamics of the cavity
field. Our strategy consists in exciting an initial coherent state
with the same photon number distribution as the micromaser
steady-state field, singling out a purely diffusive process in the
system dynamics. After the injection of a counter-field,
measurements of the population statistics of a probe atom allow us
to derive the micromaser linewidth. Our proposal aims at solving a
classic and relevant decoherence problem in cavity quantum
electrodynamics, allowing to establish experimentally the
distinctive features appearing in the micromaser spectrum due to
the discreteness of the electromagnetic field.
\end{abstract}


\vskip2pc]

The micromaser~\cite{Meschede} has been the object of continuous
interest as a proper tool for the investigation of fundamental
questions in cavity quantum electrodynamics. For example, it has
permitted the generation of highly pure Fock states~\cite{Varcoe},
the detection of photon number trapping states~\cite{Weidinger},
and the possibility of testing new ideas in the physics of quantum
information~\cite{Haroche}. Nevertheless, some fundamental aspects
have remained elusive to the experiments, as the measurement of
the micromaser spectrum and its linewidth, whose physical origin
is the decoherence of the cavity field induced by a phase
diffusion process. Although different proposals have been done in
the past, measurements related to the micromaser spectrum remain
until now as an experimental challenge for reasons we discuss
thoroughly in this work.

In Refs.~\cite{Scully1,Quang,Vogel}, approximated analytical
expressions for the phase diffusion rate were derived from the
micromaser master equation (MME)~\cite{Filipowicz}, matching
reasonably the numerical results. When the micromaser operates at
very low temperatures, trapping states of the cavity field
occur~\cite{Weidinger,Meystre}, inducing sharp minima in the mean
photon number. In consequence, as the pumping rate increases,
linewidth oscillations are predicted~\cite{Scully1}, at strong
variance with the monotonical dependence of the Schawlow-Townes
laser linewidth~\cite{ScullyBook}. In Refs.~\cite{Scully1,Brecha}
a Ramsey-type interferometric scheme was proposed for an
experimental measurement of the micromaser linewidth. These
techniques require atoms in a coherent superposition of the ground
and excited states, which needs further advances in current
technology of very high-Q (closed) cavities. Other treatments
approached the spectrum problem by investigating the field and
atomic correlation functions~\cite{Meystre,Englert}.
Unfortunately, in one way or another, all proposals have failed in
approximating the requirements of a feasible experiment.

In this Letter, we consider a realistic measurement of the phase
diffusion dynamics of the micromaser and propose different
strategies for unveiling its special features.

A micromaser consists in a single quantized mode of a high-$Q$
cavity driven by excited two-level atoms crossing the field mode
one at a time with pumping rate $r$. The atomic levels are long
living Rydberg states that couple strongly to the cavity microwave
field. The cavity is in contact with a thermal bath producing mean
number of thermal photons $\bar{n}_b$ and has a linewidth $\gamma
= \omega / Q$, where $\omega$ is the angular frequency of the
cavity field. However, the decay of the field can be considered as
negligible when an atom is flying through, due to its high speed.
Each one of the pumping atoms interacts with the cavity field
following the Jaynes-Cummings (JC) model~\cite{Jaynes}, where the
evolution of an initial atomic excited state $| {\rm e} \rangle$
and any field Fock state $| n \rangle$ follows
\begin{eqnarray}
&& | {\rm e} \rangle | n \rangle \rightarrow \cos(g \tau \sqrt{n +
1}) | {\rm e} \rangle | n \rangle - \sin(g \tau \sqrt{n + 1}) |
{\rm g} \rangle | n + 1 \rangle . \nonumber \\ && \label{JC}
\end{eqnarray}
Here, $g$ is the atom-field coupling strength, $\tau$ is the
interaction time and $| {\rm g} \rangle$ is the atomic ground
state. The temporal evolution of the density matrix elements
$\rho_{n,m} (t)$ of the cavity field is ruled by the
MME~\cite{Filipowicz}
\begin{eqnarray}
&& \frac{d}{dt}{\rho}_{n,m} = {\cal A}_{n,m} {\rho}_{n-1,m-1} +
{\cal B}_{n,m} {\rho}_{n,m} + {\cal C}_{n,m} {\rho}_{n+1,m+1}
\nonumber
\\ && \label{MME}
\end{eqnarray}
where
\begin{eqnarray}
&& {\cal A}_{n,m} = r \sin( g \tau \sqrt{n}) \sin( g \tau
\sqrt{m})
+\gamma \bar{n}_b \sqrt{n m} \nonumber \\
&& {\cal B}_{n,m} = - r \lbrack 1- \cos( g \tau \sqrt{n+1}) \cos(
g
\tau \sqrt{m+1}) \rbrack \nonumber \\
&& \,\,\,\,\,\,\,\,\,\,\,\,\,\,\,\,\,\,\,\,  - \frac{\gamma}{2}
(\bar{n}_b + 1) (n + m) - \frac{\gamma}{2} \bar{n}_b (n + m + 2)
\nonumber \\
&& {\cal C}_{n,m} = \gamma ( \bar{n}_b + 1) \sqrt{(n + 1) (m + 1)}
. \label{coeff}
\end{eqnarray}
Here, the first term of ${\cal A}_{n,m}$ and ${\cal B}_{n,m}$ are
related to the unitary contribution of the Jaynes-Cummings
interaction, after tracing out the atomic degrees of freedom, and
the remaining terms are associated with the incoherent processes.

\vspace*{-0.1cm}

\begin{figure}[htb]
\begin{center}
\centerline{\leavevmode \epsfxsize=6cm\epsfbox{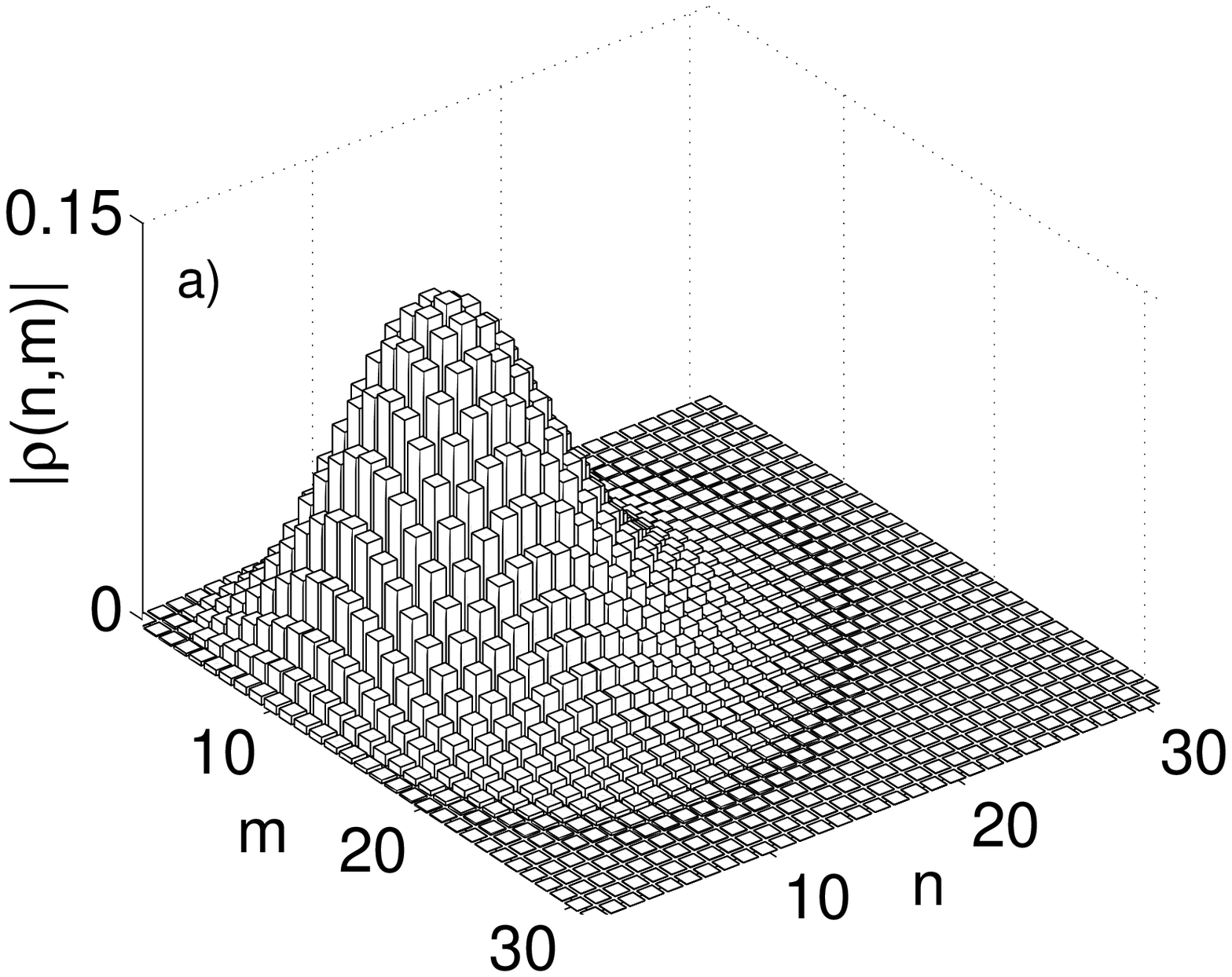}}

\vspace*{-0.1cm}

\centerline{ \leavevmode \epsfxsize=6cm\epsfbox{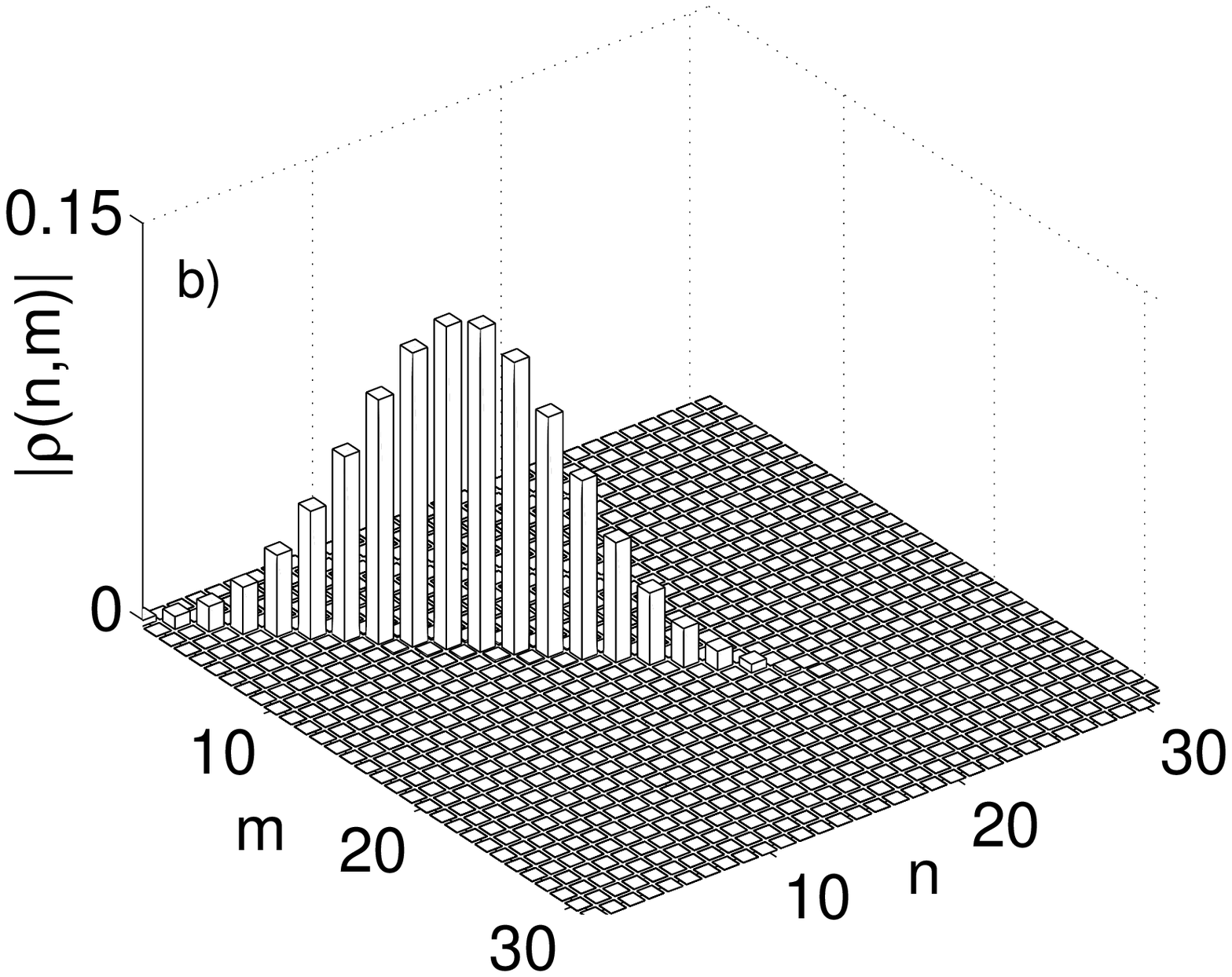}}
\vspace*{0.3cm} \caption{Density matrix $\rho_{n,m}$ of the
micromaser cavity field: a) initial coherent state $| \alpha
\rangle$; b) diagonal steady-state field after the phase diffusion
process. Dimensionless parameters: $\alpha = 3.0654$, $g \tau =
0.494$, $g/ \gamma  = 12.300$, $r/ \gamma = 10$, $\bar{n}_b =
0.03$.} \label{fig1}
\end{center}
\end{figure}

\vspace*{-0.4cm}

In the spirit of a seminal paper on the micromaser
spectrum~\cite{Scully1}, based on an approach successfully applied
in laser theory~\cite{ScullyBook}, we consider the cavity field
expectation value
\begin{eqnarray}
\langle \hat{E} (t) \rangle \propto {\rm Re} \langle \hat{a} (t)
\rangle = \frac{1}{2} \sum_{n=0}^{\infty} \sqrt{n+1} \rho_{n,n+1}
(t) + {\rm c. c.} \label{offdiagonal}
\end{eqnarray}
This expression shows that the dynamics of the electric field is
ruled by the decay of the (first) off-diagonal elements of the
field density matrix, determined by the MME of Eq.~(\ref{MME})
with $m = n + 1$. The decay of the elements $\rho_{n,n+1} (t)$,
while the photon number distribution remains stationary, implies
the decay of the cavity field expectation value $\langle \hat{E}
(t) \rangle$ by a phase diffusion process. The physical picture
becomes remarkably simple if approximately all elements
$\rho_{n,n+1} (t)$ decay exponentially with a similar rate, say
$D$. Then,
\begin{eqnarray}
\langle \hat{E} (t) \rangle =  \langle \hat{E} (0) \rangle
\exp{(-Dt)}  \label{phasediffusion}
\end{eqnarray}
and the micromaser spectrum, i.e., the Fourier transform of
$\langle \hat{E} (t) \rangle$, has a Lorentzian profile with
linewidth $D$. Eqs.~(\ref{offdiagonal}) and (\ref{phasediffusion})
are at the heart of the phase diffusion model yielding the
micromaser spectrum. In principle, it could be enough to monitor
the decay of initial nonvanishing off-diagonal elements
$\rho_{n,n+1} (0)$ of the cavity field, created by injecting atoms
in a coherent superposition of $| {\rm g} \rangle$ and $| {\rm e}
\rangle$ or preparing the cavity mode in a coherent state.
Unfortunately, this procedure cannot be implemented, as long as we
do not have direct experimental access to the cavity field. One of
the features that makes the micromaser such an interesting device
is that the pumping atoms that are used to build the micromaser
field serve also, or could serve, as a quantum probe of the field.

The first step of our scheme is the choice of a suitable initial
condition for the cavity field. The micromaser field approaches a
stationary state, independent of the initial one, described by a
diagonal density matrix with elements $\rho^{ss}_{n,n} =
p_{n}^{ss}$~\cite{Filipowicz}. If the cavity field is initially
prepared in a coherent state $| \alpha \rangle$, such that
$\rho^{\alpha}_{n,n} (0) = p_{n}^{\alpha} = |\langle \alpha | n
\rangle |^2 \cong p_{n}^{ss}$, then, for all practical purposes,
the photonstatistics remains frozen during the micromaser
operation. We choose this special initial condition in an effort
for singling out the decay of the off-diagonal elements due to
phase diffusion from other incoherent processes, allowing to
establish a non dissipative decoherence dynamics. This requirement
can be satisfied when the steady-state photonstatistics
$p_{n}^{ss}$ has a well defined peak, as in the micromaser
threshold region, where the maximum amplification of the mean
photon number occurs, or to a good approximation in the
neighborhood of the alternating trapping state region.

We simulate numerically the system dynamics by a Monte Carlo Wave
Function technique~\cite{Molmer}, successfully applied to the
investigation of other problems in the
micromaser~\cite{Casagrande2}. In Fig. 1 we show the field density
matrix $\rho_{n,m}(0)$ of the initial coherent state and also the
diagonal form of the steady-state at the end of the diffusion
process with $\rho_{n,m}^{ss} = \delta_{n,m} p^{ss}_n$, after the
progressive vanishing of field coherences. We used typical
experimental parameters and an interaction time, $\tau$, such that
the micromaser operates near the maximum amplification region.

The expected exponential decay of $\langle \hat{E} (t) \rangle$,
Eq.~(\ref{phasediffusion}), induced by the decay of the coherences
$\rho_{n,n+1}(t)$, Eq.~(\ref{offdiagonal}), is well reproduced in
the numerical simulations as shown in Fig. 2. Consequently, the
phase diffusion rate can be easily derived; in this example $D /
\gamma \cong 0.049$, in good agreement with approximated
analytical expressions~\cite{Scully1}. This behaviour, however,
can be neither directly observed, nor even inferred from
measurements on the populations of outgoing atoms. Actually, from
the JC dynamics~\cite{Jaynes}, exemplified in Eq.~(\ref{JC}), the
probability $p_{\rm e}$ that any excited atom, at any moment of
the diffusion process, is observed out of the cavity still in the
excited state is
\begin{eqnarray}
p_{\rm e} \! = \!\! \sum_{n=0}^{\infty} p^{ss}_n \cos^2 ( g \tau
\sqrt{n + 1} ) . \label{populationmeasurement}
\end{eqnarray}
In conclusion, no information on field coherences can be derived
from this kind of atomic population measurements, even if the
field coherences are nonvanishing and evolving in time. In order
to extract the required information, we suggest a strategy based
on the following steps during the diffusion dynamics:  i) to
interrupt the flux of the pump atoms at any time $t$ in the
transient regime; ii) to apply a "counter-field" $| - \alpha
\rangle$; iii) to interrogate the back-shifted cavity field by a
single probe atom.

\vspace*{-0.2cm}

\begin{figure}[htb]
\begin{center}
\centerline{\leavevmode \epsfxsize=5.5cm\epsfbox{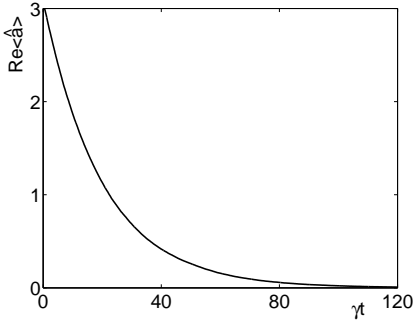}}
\vspace*{0.2cm} \caption{Exponential decay of the expectation
value of the cavity field due to phase diffusion ${\rm Re} \langle
\hat{a} \rangle$  vs. dimensionless time $\gamma t$ from the MME
(same parameters as in Fig. 1).} \label{fig2}
\end{center}
\end{figure}

\vspace*{-0.7cm}

The application of the counter-field is described by the unitary
displacement operator~\cite{ScullyBook}, $\hat{\cal{D}} (- \alpha
) = \hat{\cal{D}}^{\dagger} ( \alpha )$, mixing diagonal and
off-diagonal elements of the diffused cavity field. Then, the new
photonstatistics as a function of the diffusion time $t$ reads
\begin{eqnarray}
\tilde{p}_n(t) = \tilde{\rho}_{n,n} (t) && = \langle n |
\hat{\cal{D}} (- \alpha ) \rho (t) \hat{\cal{D}} ( \alpha ) | n
\rangle \nonumber \\
&& = \sum_{i,j = 0}^{\infty} \langle n | \hat{\cal{D}} (- \alpha )
| i \rangle \langle j | \hat{\cal{D}} ( \alpha ) | n \rangle
\rho_{i,j} (t) . \label{usefulphotonstatistics}
\end{eqnarray}
When $t = 0$, the photonstatistics $\tilde{p}_n (0)$ is simply the
one associated with the vacuum state, while, for arbitrary $t$,
$\tilde{p}_n (t)$ can be obtained numerically~\cite{Casagrande1}.
Now, an excited probe atom is injected in the cavity, and the
probability of finding it in the excited state after a transit
time $\tau_{\rm p}$ is
\begin{eqnarray}
\tilde{p}_{\rm e}(t, \tau_{\rm p}) = \sum_{n=0}^{\infty}
\tilde{p}_n (t) \cos^2 (g \tau_{\rm p} \sqrt{n+1}) .
\label{populationusefulmeasurement}
\end{eqnarray}

Now we are left with a kind of inverse problem in cavity quantum
electrodynamics, namely, to extract from the measured atomic
statistics - our {\it scattering data} - the information on the
decay of coherences $\rho_{n,n+1} (t)$ which is hidden in
$\tilde{p}_n (t)$. We indicate two solutions to this problem. The
first one exploits the time dependence of the photonstatistics
$\tilde{p}_n (t)$. Eq.~(\ref{usefulphotonstatistics}) shows that
$\tilde{p}_n (t)$ reflects the time dependence of the unshifted
field density matrix, $\rho_{i,j} (t)$. If the probe atom is
injected in the cavity at times $t \geq D^{-1}$, all coherences
$\rho_{i,i+k} (t)$ for $k \geq 2$ have already vanished. Hence,
the only contributions to $\tilde{p}_n (t)$ in
Eq.~(\ref{usefulphotonstatistics}) are given by the terms with $j
= i$, involving the diagonal elements $\rho_{i,i} (t) =
p_{i}^{ss}$, and the terms with $j = i \pm 1$, involving the
coherences $\rho_{i,i+1} (t) = \rho_{i,i+1} (0) \exp(-Dt)$ and
their complex conjugates. Hence, for times $t \geq D^{-1}$, we can
rewrite the atomic probability of
Eq.~(\ref{populationusefulmeasurement}) as
\begin{eqnarray}
\tilde{p}_e (t, \tau_{\rm p}) = K_{\tau_{\rm p}} \exp(-Dt) +
\tilde{p}_e (\infty, \tau_{\rm p} ) \label{atomicpopulationfit}
\end{eqnarray}
where
\begin{eqnarray}
K_{\tau_{\rm p}} = 2Re[\sum_{n=0}^{\infty} \sum_{i=0}^{\infty} &&
\rho_{i,i+1} (0) \langle n | \hat{\cal{D}} (- \alpha) | i \rangle
\langle i +1 | \hat{\cal{D}} ( \alpha ) | n \rangle \nonumber \\
&& \times \cos^2(g \tau_{\rm p} \sqrt{n+1} )] \label{BigKappa}
\end{eqnarray}
and
\begin{eqnarray}
&& \tilde{p}_e (\infty, \tau_{\rm p}) \! = \! \sum_{n=0}^{\infty}
\sum_{i=0}^{\infty} p_{i}^{ss} |\langle n | \hat{\cal{D}} (-
\alpha) | i \rangle|^2 \cos^2(g \tau_{\rm p} \sqrt{n+1} ) .
\nonumber \\ && \label{pinfinity}
\end{eqnarray}
Eq.~(\ref{atomicpopulationfit}) shows that $\tilde{p}_e (t,
\tau_{\rm p})$ as a function of $t$ is ruled by the phase
diffusion rate $D$, providing a simple and direct link between the
experimental data and the micromaser linewidth. Remark that
$K_{\tau_{\rm p}}$ and $\tilde{p}_e (\infty, \tau_{\rm p})$ can be
calculated directly from the initial conditions and experimental
parameters. In Fig. 3, we show $\tilde{p}_e (t, \tau_{\rm p})$ as
obtained from the numerical simulations of the whole scheme, using
the MME~(\ref{MME}), with $\tau_{\rm p} = \tau$. In this example,
$\tilde{p}_e (t)$ shows a minimum that is followed, for $t \geq
D^{-1}$, by an exponential approach to the asymptotic value
$\tilde{p}_e (\infty, \tau)$. Actually, in Fig. 3 we also show
that, as predicted by Eq.~(\ref{atomicpopulationfit}), the
behavior of $\tilde{p}_e (t, \tau)$ for times $t \geq D^{-1}$ is
perfectly fitted by an exponential law. This allows us to derive
the phase diffusion rate $D / \gamma \cong 0.047$ , quite close to
the value we calculated from Fig. 2.

\begin{figure}[htb]
\begin{center}
\centerline{\leavevmode
\hspace*{0cm}{\epsfxsize=5.5cm\epsfbox{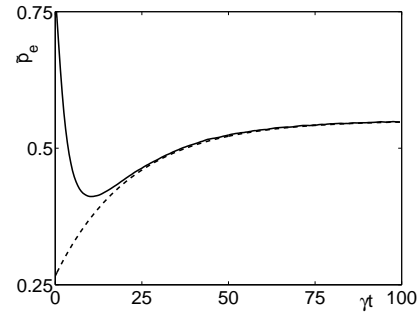}}} \vspace*{0.3cm}
\caption{Solid line: probability $\tilde{p}_{\rm e}$, from
Eq.~(\ref{populationusefulmeasurement}), that an excited probe
atom is still excited out of the cavity vs. $\gamma t$. Dashed
line: fit from Eq.~(\ref{atomicpopulationfit}) for $t > D^{-1}$,
with $K_{\tau_{\rm p}} = -0.283$, $\tilde{p}_e (\infty, \tau_{\rm
p}) = 0.551$, and $D/ \gamma =0.047$. Same parameters as in
previous figures.} \label{fig3}
\end{center}
\end{figure}

\vspace*{-0.7cm}

Another solution to our problem exploits the fact that the
interaction time $\tau_{\rm p}$ of the probe atom, injected in the
final stage of the measurement, may be different (in fact shorter)
than the interaction time $\tau$ of the pump atoms. The crucial
point is provided by the expression of the mean photon number of
the back-shifted field
\begin{eqnarray}
\langle \widetilde{\hat{N}} (t) \rangle && = {\rm Tr} \lbrack
\hat{a}^{\dagger} \hat{a} \hat{\cal{D}} (-\alpha) \rho (t)
\hat{\cal{D}} (
\alpha) \rbrack \nonumber \\
&& = {\rm Tr} \lbrack (\hat{a}^{\dagger} - \alpha^* ) (\hat{a} -
\alpha) \rho (t) \rbrack \nonumber \\
&& = 2 |\alpha|^2 \lbrack 1 - \exp (-D t) \rbrack .
\label{usefulphotonmeanvalue}
\end{eqnarray}
Here, we have used the properties of the trace and of the
displacement operator $\hat{\cal{D}}$, with the implications of
our starting assumptions $\langle \hat{N} (t) \rangle = \langle
\hat{N} (0) \rangle = |\alpha|^2$; $\langle \hat{a} (t) \rangle =
\langle \hat{a} (0) \rangle \exp(-D t) = \alpha \exp (-D t)$.
Eq.~(\ref{usefulphotonmeanvalue}) shows that the mean photon
number of the back-shifted field grows up exponentially just with
the phase diffusion rate $D$. We observe that, in principle, any
other displacement $\hat{\cal{D}} (\beta)$ would lead to the same
exponential time dependence as in
Eq.~(\ref{usefulphotonmeanvalue}). Our choice, $\beta = - \alpha$,
dictated by simplicity in the experimental implementation, here
implies also a remarkably simple mathematical expression. The
remaining step consists in finding a manifest dependence of the
measured atomic statistics,
Eq.~(\ref{populationusefulmeasurement}), on the mean photon number
of Eq.~(\ref{usefulphotonmeanvalue}). We choose an interaction
time of the probe atom such that all relevant Rabi frequencies are
small enough, i.e., $2 g \tau_{\rm p} \sqrt{n + 1} \ll 1$ for any
$n$ such that $\tilde{p}_n (t)$ is appreciable throughout the
diffusion process. In this case, the probability for the outgoing
probe atom to be in the ground state, $\tilde{p}_{\rm g}(t,
\tau_{\rm p}) = 1 - \tilde{p}_{\rm e}(t, \tau_{\rm p})$, can be
well approximated by
\begin{eqnarray}
\tilde{p}_{\rm g}(t, \tau_{\rm p}) \sim (g \tau_{\rm p} )^2
(\langle \widetilde{\hat{N}} \rangle (t) + 1 ) .
\label{finalusefulatomicpopulation}
\end{eqnarray}
Replacing $\langle \widetilde{\hat{N}} \rangle (t)$ of
Eq.~(\ref{usefulphotonmeanvalue}) in
Eq.~(\ref{finalusefulatomicpopulation}) provides another simple
and direct link between measured atomic populations and the phase
diffusion rate. For example, using the same micromaser parameters
as in Figs. 1 and 2, the result of
Eq.~(\ref{finalusefulatomicpopulation}) turns out to
hold~\cite{Casagrande1} for probe atom interaction times
$\tau_{\rm p}$ on the range $0 < g \tau_{\rm p} < 0.15$. This
means that the velocity of the probe atom should be greater than
the velocity previously chosen for the micromaser operation ($g
\tau \sim 0.5$), so that the atomic statistics exhibit an
exponential behavior that will allow to monitor the phase
diffusion dynamics. We remark that the criterion based on
Eq.~(\ref{finalusefulatomicpopulation}) holds even in the initial
stage of the diffusion process, $t < D^{-1}$, where our first
criterion failed.

We have presented a scheme for the measurement of the micromaser
phase diffusion dynamics and its associated decay rate $D$, that
is, the micromaser linewidth. We have shown, following two
different strategies, how this dynamics can be obtained from
measurable statistics of probe atoms. An experimental
implementation of this scheme is currently planned. We expect that
the proposed scheme helps to unveil experimentally the distinctive
properties of the micromaser spectrum and phase diffusion
dynamics, when compared with the conventional Schawlow-Townes
laser linewidth, due to the discrete character of the quantized
electromagnetic field.

E. S. is thankful to G. Marchi, T. Becker and B.-G. Englert for
valuable discussions.

\end{document}